\documentclass[11pt]{article}
\usepackage{blois,epsfig}
\pdfoutput=1

\def\Journal#1#2#3#4{{#1} {\bf #2}, #3 (#4)}

\def\PRD{{\em Phys. Rev.} D}

\def\mco{\multicolumn}

\def\ra{\rightarrow}

\def\ko{K^0}

  \def\beq{\begin{equation}}
  \def\eeq{\end{equation}}
\def\bea{\begin{eqnarray}}
\def\eea{\end{eqnarray}}

\newcommand{\bce}[0]{\begin{center}} \newcommand{\ece}[0]{\end{center}}
\newcommand{\ben}[0]{\begin{enumerate}} \newcommand{\een}[0]{\end{enumerate}}
\newcommand{\bit}[0]{\begin{itemize}} \newcommand{\eit}[0]{\end{itemize}}

\newcommand{\simlt}[0]{{\lower.5ex\hbox{$\; \buildrel < \over \sim \;$}}}
\newcommand{\simgt}[0]{{\lower.5ex\hbox{$\; \buildrel > \over \sim \;$}}}

\newcommand{\eg}{{\it e.g. }} \newcommand{\ie}{{\it i.e.\ }}

 \newcommand{\plancks}{{\sc Planck}
}\newcommand{\ps}{{\sc Planck}}

 \newcommand{\hfis}{{\sc HFI} }
\newcommand{\lfi}{{\sc LFI}}

\newcommand{\map}{{\sc WMAP}} \newcommand{\maps}{{\sc WMAP} }

\begin{document}
\vspace*{4cm}
\title{The \ps\ Satellite: Status \& Perspectives}

\author{ F. R. Bouchet }

\address{Institut d'Astrophysique de Paris, \\
UMR7095, CNRS and Université Pierre \& Marie Curie-Paris 6,\\
98 bis Boulevard Arago, F-75014, Paris, France.}

%==============================================================================
\maketitle\abstracts{
%==============================================================================
\ps\ was successfully launched on May 14th, 2009, from the Kourou space port, in
French Guyana. After recalling the objectives that we set out - back
in 1996 - to fulfill with this project, I recall some of the
technological breakthroughs which needed to be made and report on the
exciting scientific outlook of the project in light of the knowledge
we now have of the actual performances of the two on-board
instruments. I also include one of our more recent results even
though it was not yet available at the time of the conference. 
%
%Start w. goal/concept.\\
%put them in perspective wrt WMAP.\\
%recall needed breakthroughs.\\
%Needed to be tested, ie overview of calendar and csl results (bolo
%sens.)\\ abit about launch, orbit, colling.\\
%recall data processing prinicple.\\
%Reminder broad science case w. a couple from CMB.\\
%Conclusions\\
%NB: pre-luaunch papers.
}

%==============================================================================
%\section{Goals}
%==============================================================================

The measurement goals of \plancks may be stated rather simply: to build an
experiment able to perform the ``ultimate'' measurement of the primary
CMB temperature anisotropies, which requires:\begin{itemize}
\item full sky coverage and a good enough angular resolution in order
  to completely mine all scales at which the Cosmic Microwave
  background (CMB) primary anisotropies contain information 
  ($\simgt 5$ minutes of arc)
\item a final sensitivity essentially limited by the ability to remove
  the astrophysical foregrounds, implying a large frequency
  coverage from 30 GHz to 1 THz (provided by the two instruments: 
  HFI and LFI), with sensitivity at each of the 9 survey frequencies
  in line with the role of each map in determining the CMB properties .
\end {itemize}
For the measurement of the polarisation of the CMB anisotropies,
\plancks goal was ``only'' to get the best polarisation performances
with the technology available at the design time
%, making it more of a pioneering endeavour
%before a potential polarisation-dedicated experiment. 

This is on these simple but ambitious goals and the proposed way of
reaching them that, after 3 years of preparatory work, the project was
selected by the European Space Agency (ESA), as the $3^{rd}$ Medium
size mission of 
its Horizon 2000+ program. This selection occurred in march 1996, \ie
contemporaneously with that of WMAP by NASA, which rather proposed reaching
earlier less ambitious goals with already existing technology. 

%++++++++++++++++++++++++++++++++++++++++++++++++++++++++++++++++++++++++++++++
% Tables of Planck goal performances
%++++++++++++++++++++++++++++++++++++++++++++++++++++++++++++++++++++++++++++++
\newcommand{\hf}{\hfill}
\begin{table}[htb]
\caption{Summary of \plancks performance goals for the required 14
  months of routine operations, which allows nearly all detectors to map the
  entire sky twice. Requirements on sensitivity are simply two times
  worse than the stated goals. Central band 
  frequencies, $\nu$, are in Gigahertz, the FWHM angular sizes are in
  arc minute, and the (sky-averaged) sensitivities 
%, $\Delta X_{pix}$, 
%  are in $\mu \mathrm{K}$ per $\theta_{FWHM} \times \theta_{FWHM}$ square
%  pixels; the implied 
%are given as noise spectrum normalisation 
$c^X_{noise}$, 
%($ = \Delta X (\Omega_{FWHM})^{1/2}$), 
with $X = T, Q$ or $U$, are expressed in $\mu \mathrm{K.deg}$; this
number indicates the rms detector noise, expressed as a equivalent
temperature fluctuation in $\mu \mathrm{K}$, whcih is expected once it
is averaged in a pixel of 1 degree of linear size).}  
\vspace{0.4cm} \begin{center}
\scalebox{1.}{
\begin{tabular}{|c||c|c|c||c|c|c|c|c|c|}
\hline \hline & \multicolumn{3}{|c||}{\lfi\ goals}&\multicolumn{6}{|c|}{\hfis
goals} \\ \hline\hline $\nu$ \hf [GHz] & \hf 30 & \hf 44 & \hf 70 & \hf 100 & \hf 143 &
\hf 217 & \hf 353 & \hf 545 & \hf 857 \\ FWHM \hf [arcmin] & \hf 33 & \hf 24 & \hf 14 & \hf
9.5 & \hf 7.1 & \hf 5.0 & \hf 5.0 & \hf 5.0 & \hf 5.0 \\ 
%$\Delta T_{pix}$ & \hf 5.5& \hf 7.4 & \hf 12.8 & \hf 6.8 & \hf 6.0 &
%\hf 13.1 & \hf 40.1 & \hf 401 & \hf - \\ 
$c^T_{noise}$ \hf [$\mu \mathrm{K.deg}$]& \hf 3.0 & \hf 3.0 & \hf 3.0 & \hf 1.1 & \hf 1.4 & \hf
2.2 & \hf 6.8 & \hf - & \hf - \\ 
%\hline $\Delta Q_{pix}\, \&\, \Delta U_{pix}$ & \hf 2.8 & \hf 3.9 &
%\hf 6.7 & \hf 4.0 & \hf 4.2 & \hf 9.8 & \hf 29.8 & \hf &
%\hf \vspace{0.05\baselineskip} \\ 
$c^{Q or U}_{noise}$ \hf [$\mu \mathrm{K.deg}$]& \hf 4.5 & \hf 4.6 &
\hf 4.6 & \hf 1.8 & \hf 1.4 & \hf 2.4 & \hf 7.3 & \hf & \hf \\ \hline
\end{tabular} }
\end{center}
\label{tab:perfs}
\end{table}
%++++++++++++++++++++++++++++++++++++++++++++++++++++++++++++++++++++++++++++++

Table~\ref{tab:perfs} summarises the main performance
goals of \ps, expressed for instance as the average detector noise within a
square patch of 1 degree of linear size, $c_{noise}$, for the 14
months baseline duration of the mission, which would allow covering twice 
all the sky by nearly all 
the detectors. It is interesting to note that if we take the noise
performance figure for the average of the central CMB frequencies (the 100-143-217 GHz \hfis\ channels, assuming all the other channels
are devoted to 
foregrounds removal), one finds $0.5\ \mathrm{\mu K.deg}$ in
temperature and $1\ \mathrm{\mu K.deg}$ for the Q \& U Stokes
parameters. The magnitude of this step forward, if achieved, 
may be judged by comparing with the WMAP sensitivity which is given in
Table~\ref{tab:WMAPperfs}.  The aggregate sensitivity of the WMAP 60 \& 90 GHz
channels is $\sim 10.8\ \mathrm{\mu K.deg}$ in a year, which would
imply about $(10.8/.5)^2\sim$ 460 years of operations to reach
the baseline \ps\ sensitivity. In other words, the error bars
from noise in the angular power spectrum should be at least hundred times
smaller for \plancks than for \map (with an even larger difference at
smaller scales which will be much better known from \plancks\ thanks
to the twice higher angular resolution of HFI).

%++++++++++++++++++++++++++++++++++++++++++++++++++++++++++++++++++++++++++++++
% Tables of WMAP in-flight performances
%++++++++++++++++++++++++++++++++++++++++++++++++++++++++++++++++++++++++++++++
\begin{table}[htb]
\caption{Summary of WMAP in-flight performance per full year of operations. Same
  units than in Table~\ref{tab:perfs}. }
\vspace{0.4cm}
\begin{center}
{ %\footnotesize {\footnotesize
\begin{tabular}{|c||c|c|c|c|c|}
\hline \hline 
\multicolumn{6}{|c|}{\maps (in flight) } \\ 
\hline\hline 
$\nu$ \hf [GHz] & \hf 23 & \hf 33 & \hf 41 & \hf 61 & \hf 94 \\
FWHM \hf [arc min] & \hf 49.2 & \hf 37.2 & \hf 29.4 & \hf 19.8 & \hf 12.6 \\ 
$c^T_{noise}$ \hf [$\mu \mathrm{K.deg}$] & \hf 12.6 & \hf 12.9 & \hf 13.3
& \hf 15.6 & \hf 15.0 \\  
\hline
\end{tabular} } %}  
\end{center}
\label{tab:WMAPperfs}
\end{table}
%++++++++++++++++++++++++++++++++++++++++++++++++++++++++++++++++++++++++++++++

%++++++++++++++++++++++++++++++++++++++++++++++++++++++++++++++++++++++++++++++
% Plank build-up through anim extracts
%++++++++++++++++++++++++++++++++++++++++++++++++++++++++++++++++++++++++++++++
\begin{figure}[htbp] \begin{center}
\psfig{file=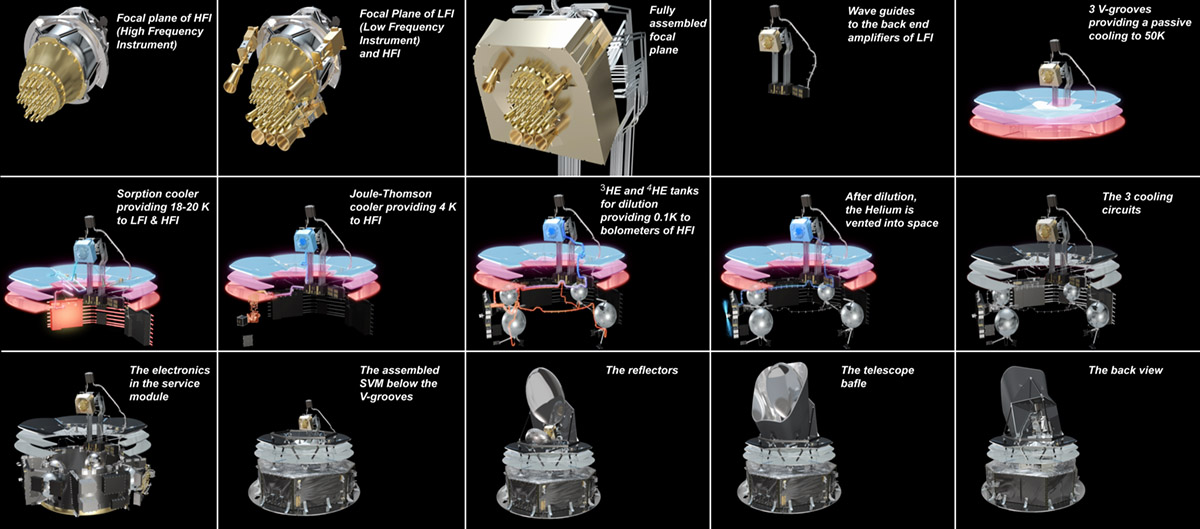, width=\textwidth} 
\end{center}
\caption[]{Planck build-up from the inside out. Going from left to
  right, one sees on the top row (t1) the HFI instrument with its 52
  detector horns poking out of it's outer shell at 4\,K. (t2) HFI is
  surrounded by the 11 larger horns from LFI. HFI and LFI together
  form (t3) the focal plane assembly from which (t4) the electrical
  signal departs (though a bunch of wave guides for the LFI 
  and a harness of wires for HFI) to connect to the warm electronics parts of
  the detection chain which are located within the service module at
  $\sim$ 300\,K. (t5) The (top) cold and warm (bottom) parts are
  separated by three thermally isolating V-grooves which allow
  radiating to space heat from the spacecraft sideways and quite
  efficiently. The third (top) V-grooves operating temperature is
  about 40\,K.
  On the middle row, one sees (m1) the beds of the sorption cooler and
  it's piping around the V-groove, bringing the overall focal-plane
  structure to LFI's operational temperature of $\sim 18$\, K. (m2) The
  back-to-back (to damp the first harmonics of the vibrations)
  compressors of the 
  4\,K cooler allow bringing HFI outer shell to 4\,K, while (m3) isotopes
  from the He3 tank and the three He4 tanks are brought to the mixing
  pipes within HFI to cool filters (within the horns) to 1.6\,K and the
  bolometer plate to 0.1\,K, before (m4) being released to space. (m5) The
  passive cooling and the three active stage constitute this
  complex but powerful cooling chain in space.
  On the last bottom row, one
  can also see (b1) some of the electronic boxes in the service module (SVM)
  which in addition to the warm part of the electronic and cooling
  chains also contain all ``services'' needed for transmitting data,
  reconstructing the spacecraft attitude, powering the whole
  satellite... The bottom of the SVM is covered with solar panels, while
  supporting struts begin on its top which allows (b3) positioning the
  secondary and primary reflectors. The top part is surrounded by a large
  baffle to shield at best the focal plane from stray-light. The back
  view (b5) allows distinguishing in the back the supporting structure
  of the primary mirror, and the wave guides from LFI. The spin axis of
  Planck (vertical on these plots) is meant to be close to the
  sun-earth line, with the solar panel near perpendicular to that line
  and the rotation of the line-of-sight (at 1 rpm) causing the detectors to
  survey circles on the sky with an opening angle around 85
  degrees. Copyright ESA.} 
\label{fig:build}
\end{figure}
%++++++++++++++++++++++++++++++++++++++++++++++++++++++++++++++++++++++++++++++

We proposed to achieve the ambitious sensitivity goals of \plancks
with a small number of detectors, limited principally by the photon noise of the
background (for the CMB 
ones), in each frequency band. This implied to achieve several
technological feats never achieved in space before (see in particular Lamarre
et al. 2003, in New Astronomy Reviews, 47, pp. 1017):
\begin{itemize}
\item sensitive \& fast bolometers with a Noise Equivalent Power $< 2 \times
  10^{-17}\ \mathrm{W/Hz}^{1/2}$ and time constants typically smaller than
  about 5 milliseconds (which thus requires cooling them down to $\sim 100$
  mK, and build them with a very low heat capacity \& charged particles
  sensitivity)
\item total power read out electronics with very low noise, $< 6\
\mathrm{nV/Hz}^{1/2}$ from 10 mHz (1 rpm) to 100 Hz (\ie from the
largest to the smallest angular scales to measure at the \ps\ scanning speed)
\item excellent temperature stability, from 10 mHz to 100 Hz (cf.
  Lamarre et al. 04), such that the induced variation be a small fraction
  of the detector temporal noise:  \begin{itemize}
    \item better than $ 10\ \mathrm{\mu K/Hz}^{1/2}$ for the 4K box
      (assuming 30\% emissivity)
    \item better than $30\ \mathrm{\mu K/Hz}^{1/2}$ on the 1.6K filter plate
    (assuming a 20\% emissivity)
    \item better than $20\ \mathrm{nK/Hz}^{1/2}$ for the detector plate (a
      damping factor $\sim 5000$ needed)
\end{itemize}
\item very low noise HEMT amplifiers (therefore cooled to 20\,K) \& very stable
  cold reference loads (at 4\,K)
\end{itemize}
In addition, \plancks requires:\begin{itemize}
\item a low emissivity telescope with very low side lobes (\ie strongly
under-illuminated)
\item no windows, and minimum warm surfaces between the detectors and
  the telescope
\item a quite complex cryogenic cooling chain
  (cf. figure~\ref{fig:build}) which begins by reaching
  $\sim 40K$ via passive cooling, by radiating about 2 Watts to space,
  followed by three active stages, at 20\,K, 4\,K, and 0.1\,K:
\begin{itemize}
  \item 20\,K for the LFI, with a large cooling power, $\sim 0.7$ Watts
  (provided by $H_2$ Joule-Thomson sorption pumps developed by JPL, USA)
  \item 4\,K, 1.6\,K and 100\,mK for the HFI (the 15 milli-Watts
  cooling power at 4K is provided by mechanical pumps provided by the
  RAL, UK, in order to perform a Joule-Thomson expansion of He; the
  1.6K stage has a pre-cooling power of about 0.5 milli-Watts, thanks to
  another Joule-Thomson expansion, while the final dilution fridge of He$^3$
  \& He$^4$, from a French collaboration between Air Liquide, the
  CRTBT, can deal with 0.2  micro-Watts at 0.1\,K).
  \item a thermal architecture optimised to damp thermal fluctuations
  (active+passive)
\end{itemize}
\end{itemize}
Furthermore, a tight control of vibrations is needed, in particular
since the dilution cooler does not tolerate micro-vibrations at
sub-mg level. And as little as $7\times 10^{10}$
He atoms accumulated on the dilution heat exchanger (an He pressure typically
at the $1\times 10^{-10}$ mb level) would be too much.

These top-level design goals have now been turned into real
instruments, which went 
through several qualification models. Before delivering the actual
flight model of both instruments to industry for 
integration with the satellite, both instrumental consortia organised extensive
calibrations campaigns starting at the individual components levels,
then at the sub-systems levels (e.g. individual photometric pixels),
then at instrument 
level. For HFI, the detector-level tests were done mainly at JPL in
the USA, and the pixel level tests were performed in Cardiff in the
UK, while the flight instrument calibration was performed at the
Institut d'Astrophysique Spatiale in Orsay, France from April till the
end of July 2006. During that period, we obtained in particular 
19 days of scientific data at normal operating conditions. We could then
confirm that HFI satisfies all our {\em requirements}, and for the most part
actually reaches or exceeds the more ambitious design {\em goals}, in particular
concerning the sensitivity, and speed of the bolometers, the very low noise of
the read out electronics and the overall thermal stability. In
addition, the total optical efficiency has been verified to be
satisfactory, optical cross-talk appears negligible, as well as the
Current cross-talk and the cross-talk in intensity is weak. Main beams
are well-defined and are quite well described by the models,
polarisation measurements confirm expectations, etc. The LFI 
instrument also went through detailed testing around the same time and
it does reach most of its ambitious requirements. 

%++++++++++++++++++++++++++++++++++++++++++++++++++++++++++++++++++++++++++++++
% Plank spacecraft image at CSL
%++++++++++++++++++++++++++++++++++++++++++++++++++++++++++++++++++++++++++++++
\begin{figure}[htbp] \begin{center}
\psfig{file=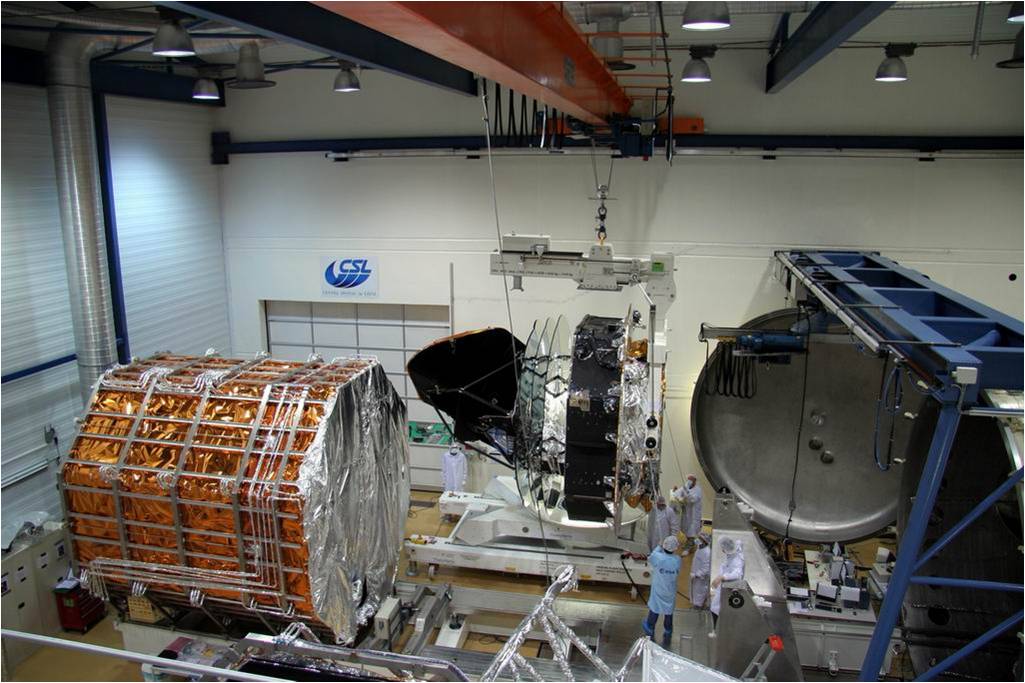, width=\textwidth} 
\end{center}
\caption[]{Planck on May 2008, hanging outside the vacuum cryogenic
  chamber at CSL before completing it's first and last full thermal
  vacuum test. Copyright ESA.}
\label{fig:CSL}
\end{figure}
%++++++++++++++++++++++++++++++++++++++++++++++++++++++++++++++++++++++++++++++

\begin{figure}\begin{center}
\psfig{figure=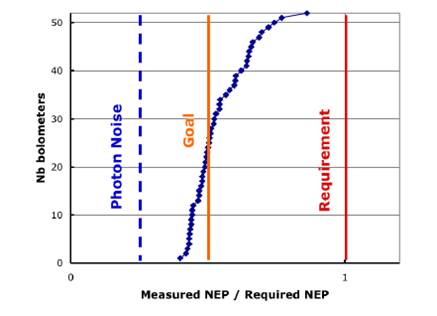,width=0.6\textwidth}
\end{center}
\caption[]{Measured values of the Noise Equivalent Power of HFI
  detectors during the ground test in flight conditions at CSL on May
  2008. One sees that the median value is at the goal level.}
\label{fig:bolos}
\end{figure}

The integration of the LFI and HFI instruments was performed at Thales
premises in Cannes 
in November 2006 and within a year, by December 2007, the full
satellite was ready for vibration testing. Planck was then flown from
Cannes to ESA's ESTEC centre (in Noordwijk, Holland) where among
other things it went through load balancing on April 7th, before
travelling again to the ``Centre Spatial de Li\'eges'' (CSL) in april
2008. Figure~\ref{fig:CSL} is a picture of Planck hanging outside the
vacuum cryogenic chamber at CSL, before the start of the first (and
last) full thermal test with all elements of the cryogenic chain
present and operating. This ultimate system-level (ground) test
demonstrated in particular the following: 
\begin{itemize}

\item the dilution system can work with the minimal Helium 3 and 4 flux,
  which should allow 30 months of survey duration (nominal duration
  being 14 months!) 

\item the extremely demanding temperature stability required (at 1/5
  of the detection noise) has been verified
%, with in
%  particular:\begin{itemize}
%    \item $20 nK/\sqrt{Hz}$ on the bolometer plate at 100mK over the
%      needed range of [10mHz, 100Hz]  
%      \item $10\mu K/\sqrt{Hz}$ on the 4K horns and filters 
%   \end{itemize}

\item bolometers sensitivities in flight conditions are indeed centered around
  the goal, as shown in figure~\ref{fig:bolos}.  
\end{itemize}

%++++++++++++++++++++++++++++++++++++++++++++++++++++++++++++++++++++++++++++++
% Bolos sensitivities
%++++++++++++++++++++++++++++++++++++++++++++++++++++++++++++++++++++++++++++++
%++++++++++++++++++++++++++++++++++++++++++++++++++++++++++++++++++++++++++++++

\begin{figure}\begin{center}
\psfig{figure=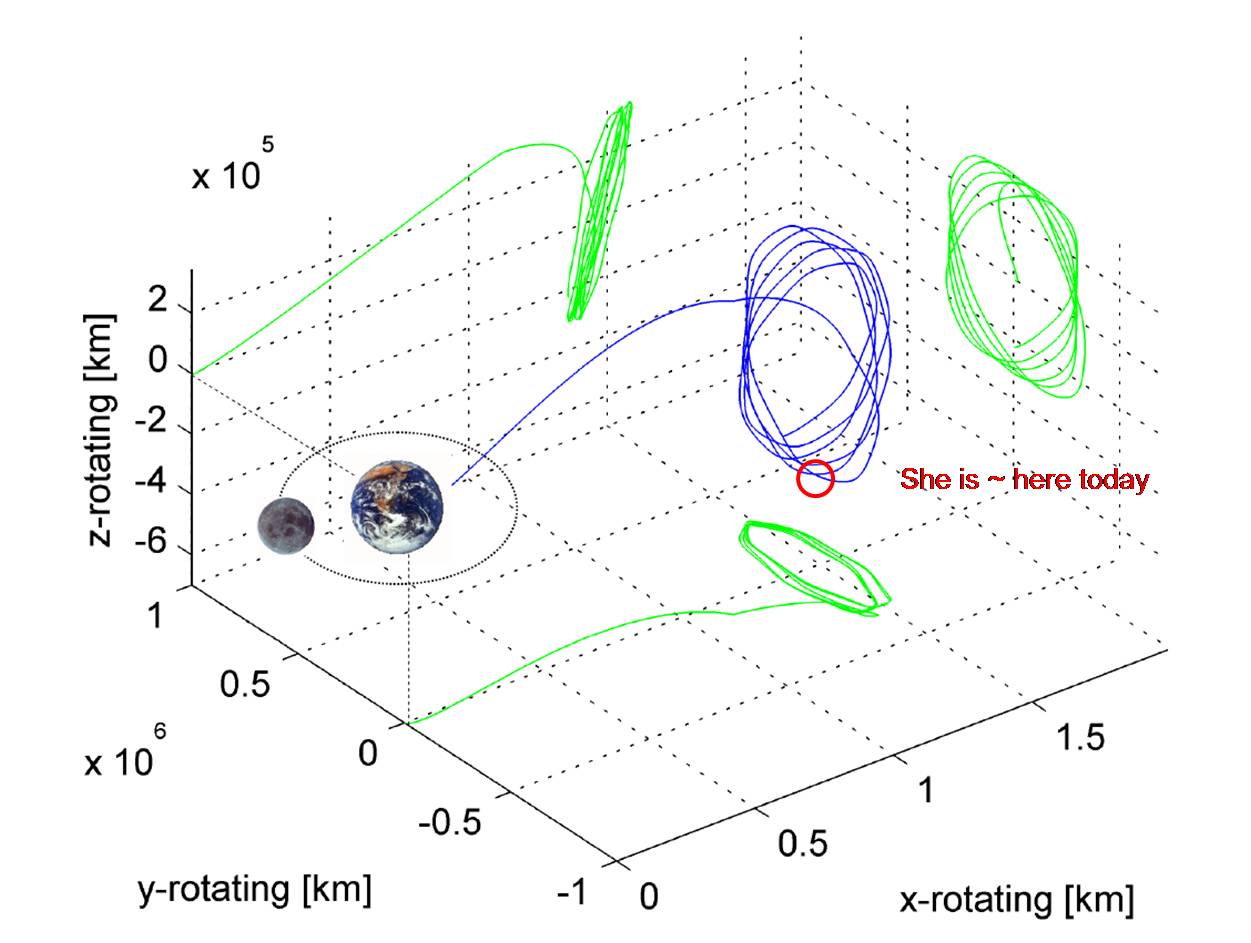,width=0.45\textwidth}
\psfig{figure=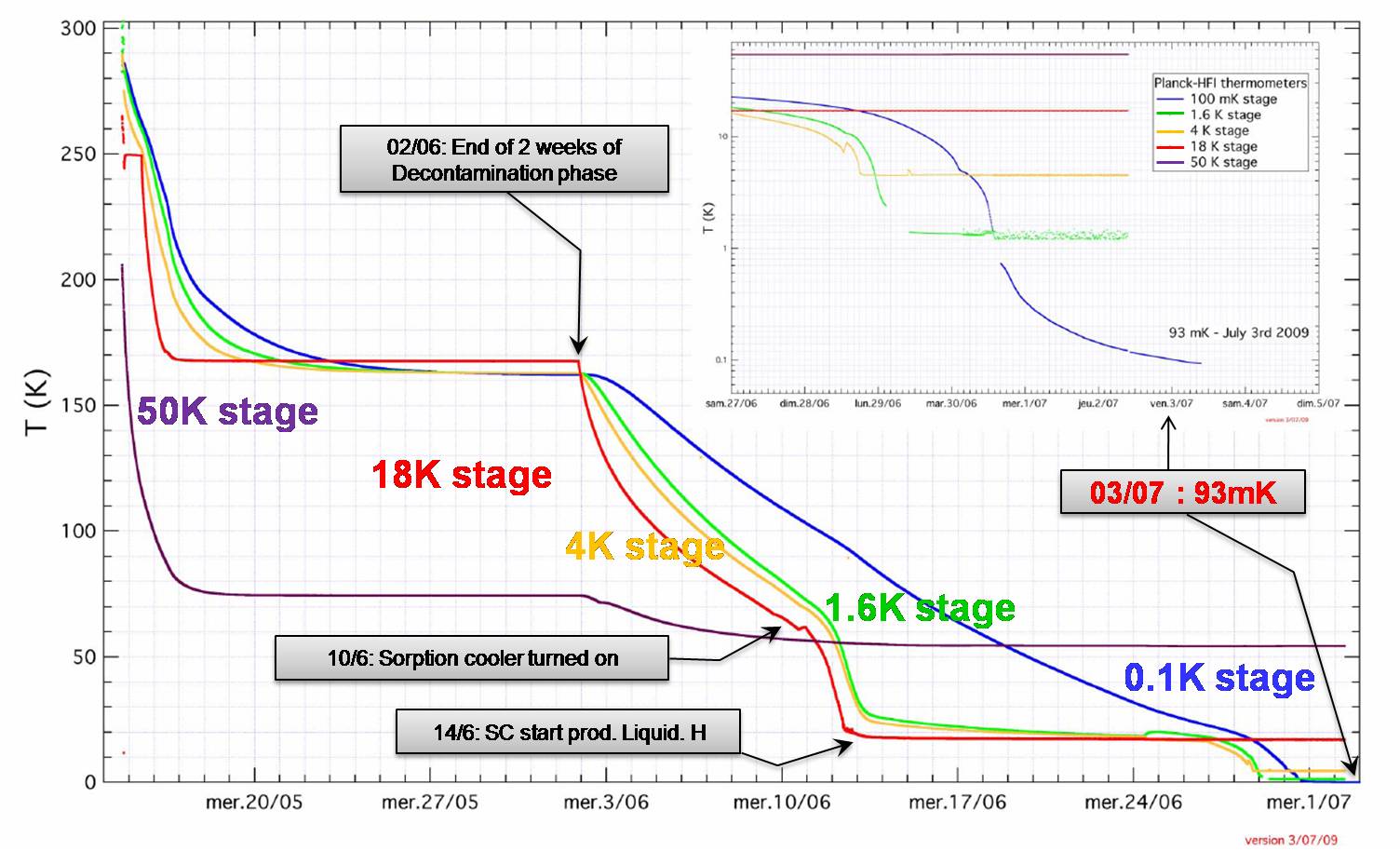,width=0.54\textwidth}
\end{center}
\caption[]{a) Spacecraft trajectory to and on the L2 orbit. b) Cooling
  sequence of Planck, showing the various stages 
  reaching in turn their operational temperature, till the dilution
  plate actually reached 93 mK on July 3rd. Credit ESA and HFI consortium}
\label{fig:trajcool}
\end{figure}

\ps\ was then shipped to Kourou, and after a few more nerve –wracking
delays, we finally lost sight of Planck for ever (when it was covered
by the SYLDA support system on the top of which laid Herschel for a joint
launch). Launch was on May 14th, and it was essentially perfect. After
separating from Herschel, \ps\ was set in rotation and started its
 to the L2 Lagrange point of the sun-earth system, at 1.5 million
kilometres away from earth, \ie about 1\% further away from the sun than the
earth. The final injection in the L2 orbit was at the end of June,
shortly after the end of the Blois meeting (see
figure~\ref{fig:trajcool}-a), at the same time than the cooling
sequence ended successfully. Indeed, figure~\ref{fig:trajcool}-b shows
how the various thermal stages reached their operating temperature,
cooling of coarse from the outside-in, and closely following the
predicted pattern. 

%++++++++++++++++++++++++++++++++++++++++++++++++++++++++++++++++++++++++++++++
% Trajectory vs L2 + cooling curve
%++++++++++++++++++++++++++++++++++++++++++++++++++++++++++++++++++++++++++++++
%++++++++++++++++++++++++++++++++++++++++++++++++++++++++++++++++++++++++++++++

\begin{figure}\begin{center}
\psfig{figure=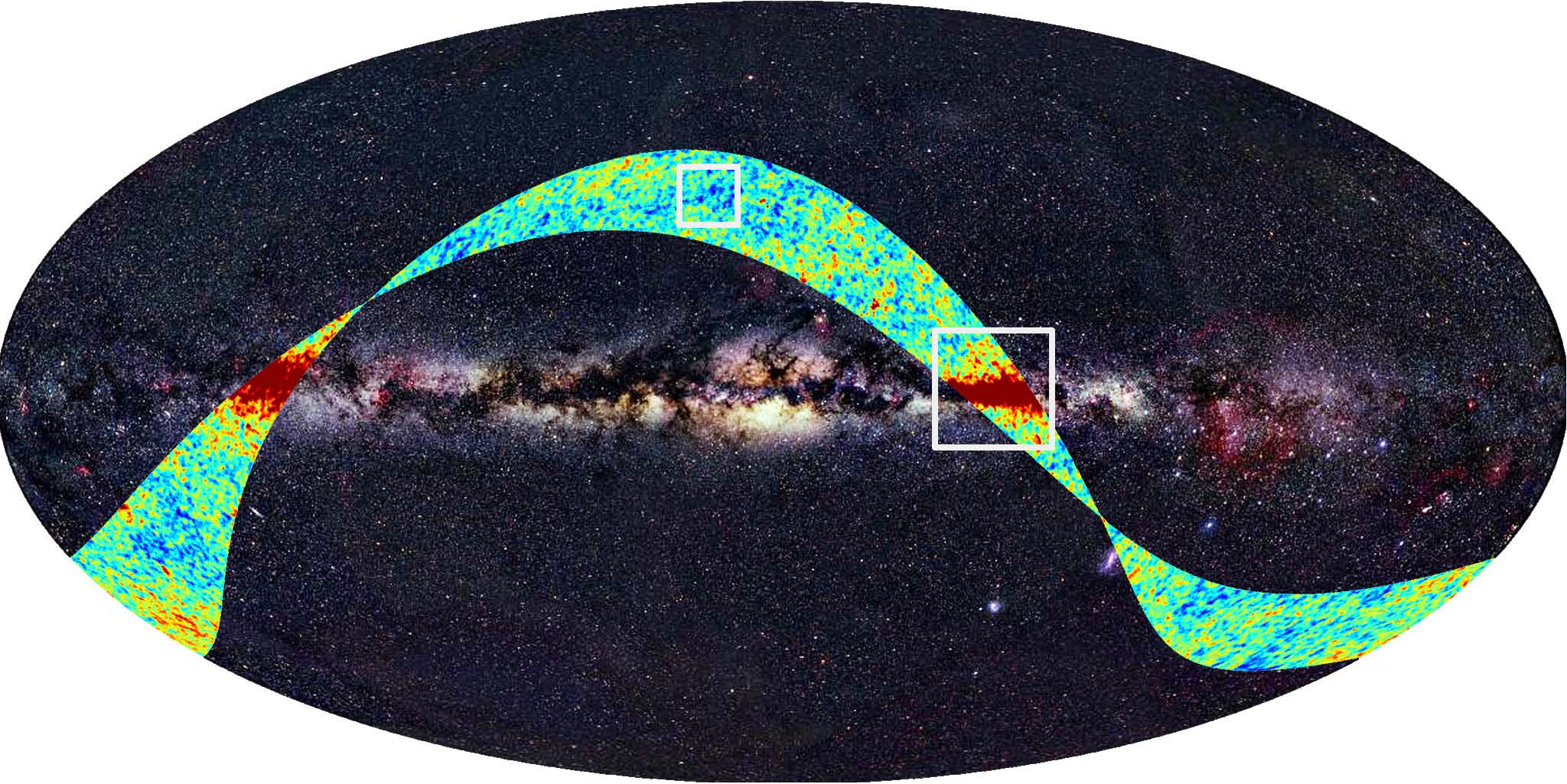,width=\textwidth}
\end{center}
\caption[]{Sky coverage during Planck First Light Survey (FLS). Planck
  measures a ring of about 1 degree per day (for a full rotation around
  the sun in a year), yielding this $\sim 15$ degree strip during the
  two weeks of the FLS. The detector image shown in Galactic
  coordinates is superimposed 
  to a background image where the Galactic plane is clearly
  visible. Credit ESA and HFI \& LFI consortia. }
\label{fig:strip}
\end{figure}

Once at L2, a calibration and performance verification phase was
conducted till mid-august, to insure that all system are working
properly and that instrumental parameters are all set at best.  
From August 13th to 27th, we conducted a ``First Light Survey'' (FLS)
in normal operational mode for an ultimate verification of parameters
and of the long-term stability of the experiment. We found the data
quality to be excellent, and the Data processing Centre pipelines
could be operated as hoped to produce the first
images. Figure~\ref{fig:strip} (extracted from the Press release we
made on September 17th) illustrates the FLS coverage by
showing an image generated from the data acquired from a single 
100GHz detector of HFI superimposed to an image of the optical
sky by Axel Mellinger. We also released (see the press release in
English at ESA's site 
http://sci.esa.int/science-e/www/object/index.cfm?fobjectid=45543 and
in French at http://public.planck.fr/actualiteFLS.php) a comparison of
a high latitude field whose 
emission ought to be dominated by the CMB (shown by a
small white square in the figure) as observed by an HFI and LFI
detector. They demonstrate an excellent similarity while the two
instruments are using quite different technologies. Nine images at all
the frequencies covered by Planck of a Galaxy crossing area (indicated
by the large white square of the figure) 
provide visual evidence to the richness of the dataset that Planck
shall deliver, allowing very broad scientific studies outreaching its
primary cosmological goals. Indeed an important part of Planck long
term legacy will be the unique set of  maps of the millimetric and
sub-millimetric polarised full sky.

%++++++++++++++++++++++++++++++++++++++++++++++++++++++++++++++++++++++++++++++
% FLS coverage
%++++++++++++++++++++++++++++++++++++++++++++++++++++++++++++++++++++++++++++++
%++++++++++++++++++++++++++++++++++++++++++++++++++++++++++++++++++++++++++++++

With the success of the FLS, the normal survey operations have now
started. We should therefore be in a position to deliver in
December 2010, as planned, an “Early Release Point Source Catalogue” based on a
rapid analysis of the first coverage of the sky which will be issued in time to 
allow the astrophysical community at large to propose follow-ups  by
Herschel during  
its expected cryogenic life. A first public release based on the data
from the nominal 14 months mission is slated for december 2012. The
release should contain the clean calibrated time-ordered data of each
detector, the nine full sky maps at the six frequencies covered by HFI
and the three ones by LFI, possibly supplemented by polarisation maps, 
as well as maps of identified astrophysical components (CMB, Galactic
Emissions, Extragalactic sources catalogue), some ancillary
information (\eg on beams, spectral transmission, etc), accompanied
by about 50 scientific papers describing the mission, how the
``products'' were obtained, validated, and the results of a first
pass of scientific exploitation by the Planck collaboration itself,
encompassing in particular the implications of the measured
statistical properties of the CMB. Our anticipation from the measured 
Helium consumption in flight is that the mission duration will exceed the
nominal duration of 14 months, and we plan a further release, about a
year later on the 
basis of the extra data which might allow covering as much as five
times in total the entire sky. In addition to an improved sensitivity,
this extra 
duration will foremost allow greater data redundancy and therefore a
tighter control of all systematic effect which can be searched for with
a longer baseline. This should allow us detecting the gravitational wave
stochastic background predicted in one interesting class of
inflationary models, providing the long thought after ``smoking
gun'' of inflation, or otherwise put meaningful constraints on the
viable inflation models remaining.  
 
{\em In conclusion, \ps\ is now in normal operation \& performances are as
expected or better.}\\
This gives us confidence that the scientific
program of \ps\ can be carried through as anticipated. The dataset should in
particular allow addressing many key cosmological questions, including
the existence of a primordial gravitational wave background, or 
that of highly revealing deviations from the current minimal model, 
where the primordial fluctuation can be purely Gaussian, adiabatic,
scale-free, in a strictly flat spatial geometry with a dark energy
component describable as a pure cosmological constant, and
(cosmologically) negligible neutrinos masses. 

A rather complete 
overview of the scientific Program of \ps\ can be found in the
so-called ``Blue Book'' which was issued in 2004. It can be downloaded 
from \\
{\em http://www.planck.fr/IMG/pdf/Planck\_book.pdf}. \\
In addition,
we submitted a series of pre-launch papers (all with a title starting
with ``Planck pre-launch status:'') giving many details of
the design and tests of the mission, the instruments, and some of
their components.

%==============================================================================
\section*{Acknowledgments}
%==============================================================================

\plancks is the result of the efforts of a large industrial and research team,
which includes a large fraction of Europe's far infrared, submillimeter and
CMB community, as well as a large number of CMB researchers from the
US. \plancks (http://www.rssd.esa.int/Planck) is a project of the European
Space Agency with instruments funded by ESA member states (in particular the
lead countries: France and Italy) with special contributions from NASA
(USA) and with the telescope reflectors provided by a collaboration
between ESA and a scientific consortium led and funded by Denmark. The
project involves about 50 participating scientific institutes. The
Planck Science team comprises Marco Bersanelli, Fran{\c c}ois R. Bouchet,
Georges Efstathiou, Jean-Michel Lamarre, Charles Lawrence, Nazzareno Mandolesi,
Hans-Ulrick Norgaard-Nielsen, Jean-Loup Puget, Jan Tauber, Andrea Zacchei.

%%%%%%%%%%%%%%%%%%%%%%%%%%%%%%%%%%%%%%%%%%%%%%%%%%%%%%%%%%%%%%%%%%%%%%%%%%%%%%%
%%%%%%%%%%%%%%%%%%%%%%%%%%%%%%%%%%%%%%%%%%%%%%%%%%%%%%%%%%%%%%%%%%%%%%%%%%%%%%%
%%%%%%%%%%%%%%%%%%%%%%%%%%%%%%%%%%%%%%%%%%%%%%%%%%%%%%%%%%%%%%%%%%%%%%%%%%%%%%%
\end{document}